\documentstyle[a4,12pt]{article}

\begin{document}

\begin{flushright}
FIAN-TD/24-97
\end{flushright}

\begin{center}
{\large {\bf Nonperturbative Violation of Factorial Moments Scaling \\
 in Dissipative QCD Jets \footnote{Work supported by Russian Foundation
 for Basic Research under Grant N 96-02-16210}}}
\end{center}

\bigskip

\begin{center}
{\bf Andrei Leonidov \footnote{{\bf e-mail address leonidov@td.lpi.ac.ru}}
and Dmitri Ostrovsky \footnote{{\bf e-mail address ostrov@td.lpi.ac.ru}}}
\end{center}

\bigskip

\begin{center}
{\it Theoretical Physics Department, P.N. Lebedev Physics Institute \\[0pt]
117924 Leninsky pr. 53, Moscow, Russia}
\end{center}

\bigskip

\begin{center}
{\bf Abstract}
\end{center}

The nonperturbative modification of the evolution equation for QCD generating
functional is applied to the analysis of the factorial moments of the intrajet
particle distributions in QCD. Nonperturbative preasymptotic correction to
the factorial moments scaling (F-scaling) is calculated. The results are
compared with the available experimental data.  The data show the possible
presence of such a nonperturbative effect.

\newpage

\section{Introduction.}

Physics of jets provides one of the most important testing grounds
for practical applications of the theory of strong interactions, QCD.
Perturbative evolution of a quark-gluon jet
is now well investigated (see, for example, \cite{BPQ}). In particular,
one of the striking predictions of perturbative QCD is an angular ordering
in the intrajet evolution due to the color coherence in gluon emission.
 Nonperturbative aspects of
jet evolution are considerably less clear. From experience of applying
the sum rules to the analysis of characteristics of the heavy resonances \cite
{SVZ} it is clear that when virtuality of radiated parton is of the order of
1~GeV it is necessary to take into account nonperturbative
corrections due to background vacuum fields giving rise to QCD
vacuum condensate. Technically this results in introducing additional soft
interaction of hard parton currents with the soft medium parametrized by the
condensate fields. A similar situation arises in describing an evolution
of a jet in some nontrivial (e.g. nuclear) medium, so that inelastic collisions
with it lead to dissipation of energy and momentum from the hard parton
component. Although physically these situations are rather different,
we expect the appropriate theoretical
formalisms describing both cases to be rather similar.

 In both cases a necessity of introducing new interaction vertices
lead to the occurrence of new dimensional parameters. This in turn results
 in the specific scaling violations
in the description of jet evolution, in which the perturbative interaction
vertices are dimensionless (contain ratios of energies or
virtualities, see \cite{BPQ}). Physically this could be interpreted as a
beginning of nonperturbative process of the QCD string formation
which takes energy from the perturbative component and pumps it into
nonperturbative degrees of freedom. The first model describing the
jet evolution taking into account the nonperturbative energy loss was proposed
by I.M.Dremin \cite{Dr}. It was based on an analogy with
the physics of high-energy electromagnetic showers in a medium, where
scaling-invariant evolution due to photon bremsstrahlung and electron-positron
pair creation is accompanied by the scaling-violating scattering due to
inelastic interactions of shower particles with the atoms of the medium
leading to their ionization \cite{Be}. Later the corresponding modified
QCD evolution equations were analytically solved in \cite{DL1}, where
expressions for parton multiplicities in quark and gluon jets and
the energy loss in the perturbative hard component were calculated.
The discussion of this method, as well as of the relevant Monte-Carlo
 calculations \cite{GE}, where nonperturbative component was described through
 the effective QCD lagrangian can be found in the review paper \cite{DL2}.
 Serious defect of Dremin's equations \cite{Dr} is that there the color
interference in intrajet evolution, known to be of crucial significance for the description of
its characteristics, is not taken into account. The problem we are
discussing in this paper is to build up a formalism providing an analytical
description of intrajet evolution and taking into account
both color coherence and nonperturbative energy loss.

This program was started in the previous work of the authors
 \cite{LO}, where the modified jet evolution equations were solved in the
leading logarithmic approximation (LLA) and the general evolution equations
for QCD generating functional taking into account the nonperturbative
dissipative corrections were written down. In \cite{LO} we considered only the
simplest characteristics of the intrajet evolution such as the nonperturbative
dissipative corrections to the intrajet particle multiplicities and
distributions. In this paper we turn to a more detailed analysis of the jet
particle content by calculating the nonperturbative dissipative corrections
to the factorial moments characterizing it.

   Let us remind that perturbative QCD predicts an asymptotic independence of
the moments of the intrajet particle distributions on energy (KNO scaling)
\cite{BPQ}. Closely related to the KNO scaling is a so-called factorial moments
scaling (F-scaling) that holds in the high energy limit.  A very
clear review on this subject can be found, e.g., in \cite{D1}, where one can
also find the relevant references (see also a recent review \cite{D2}).
The perturbative preasymptotic corrections to this scaling are due to the
running of the coupling constant and thus proportional to $\sqrt{\alpha_s(Q^2)}$
\cite{KO} (and therefore to $1/\sqrt{\ln Q^2}$). Below we shall restrict
ourselves to the case of the frozen coupling constant, so the above-mentioned
perturbative preasymptotic corrections are absent and the analysis is focused
on the nonperturbative preasymptotic F-scaling violation due to dissipative
effects in the jet's evolution.

   Let us note, that a modification of the evolution equations for QCD
generating functional similar to those in \cite{LO} and introducing an
additional inactivation of perturbative evolution were
recently described in \cite{BP} \footnote{We are grateful to I.M. Dremin for
bringing our attention to this paper}.

\section{Generating functional for dissipative evolution}

It is well known, that a formalism of generating functional \cite{BPQ} is
indispensable in analyzing the properties of particle distributions in QCD
jets. In particular, it provides a unique possibility of analytically
accounting for such delicate properties of QCD jet evolution as color
coherence and resulting angular ordering in the parton cascades. In the previous
paper \cite{LO} we have proposed a nonperturbative generalization of the
perturbative evolution equation for the QCD generating functional describing
the intrajet particle distributions. A new element is to introduce a
dimensional coupling between the perturbative partonic and nonperturbative
soft modes resulting in additional ''dissipative'' damping of the
perturbative intrajet evolution. The modified evolution equation in MLLA
(modified leading logarithmic approximation) for the generating functional $%
G(p,\theta |u)$ reads
\begin{equation}
\frac{\partial G(p,\theta |u)}{\partial \ln \theta }=\int dx\ \gamma
_{0}^{2}\ K(x)\ \left[ G(xp,\theta |u)\ G((1-x)p,\theta |u)-G(p,\theta
|u)\right] -\left(
{\displaystyle {\beta  \over p}}%
\right) ^{\alpha }p\frac{\partial G(p,\theta |u)}{\partial p}
\label{General}
\end{equation}
where $p$ is an energy of the parton decaying into the two ones having
energies $xp$ and $(1-x)p$ correspondingly, $\theta $ is an angle between
the off-spring partons and $\beta $ is a dimensional coupling constant
accounting for the coupling to the nonperturbative modes.
 In Eq.~(\ref{General}) $\gamma_{0}^{2}=2N_{c}\alpha _{s}/\pi $ is a running
 coupling, ($N_{c}=3$ is a number of colors), $K(x)=1/x-1+x(1-x)/2$ is an
 MLLA kernel for gluon
splitting. The initial condition for this equation has the form $G(p_{\min
},\theta _{\min }|u)=u$, thus fixing the borderline for the possibility of
perturbative evolution. From the requirement of probability conservation we
get another constraint for the generating functional $G(p,\theta |1)=1$.
The Eq.~(\ref{General}) is a further generalization of the one in \cite{LO},
where only the case of $\alpha =1$ was considered.

In the absence of dissipative effects (i.e. for $\beta =0$) the evolution
described by Eq.(\ref{General}) is characterized by a specific scaling
\cite{BPQ}.  Namely, the generating functional $G(p,\theta |u)$ depends only on
a certain combination of $p$ and $\theta $ having the form $y=\ln (p\theta
/Q_{0})$, where $Q_{0}=p_{min}\theta _{min}$ is a minimal transverse momentum
considered in terms of the perturbative evolution. This results, in particular,
in the fact that both KNO scaling for intrajet multiparticle distributions and
the (equivalent to it in the limit of high energy) F-scaling \cite{D1} hold
in the limit of $y \rightarrow \infty$ (not only $p \rightarrow \infty$).
However, this is no longer true when the additional ''dissipative'' term is
taken into account. The presence of
the new dimensional coupling $\beta $ leads, generally speaking, to the
explicit violation of the perturbative scaling.

 Below we shall consider only
the effects of the first order in a parameter $\varepsilon =(\beta /p)^\alpha$,
so we
assume that the following decomposition takes place:
\begin{equation}
G(p,\theta |u)=G_{0}(y|u)-\varepsilon G_{1}(y|u)  \label{resol}
\end{equation}
so that the perturbative scaling for $G_{0}(p,\theta )$ and $G_{1}(p,\theta )$
still holds.

Substituting the decomposition (\ref{resol}) into the evolution equation
Eq.~(\ref{General}), we get in the zeroth and first order in $\varepsilon $
\begin{equation}
\frac{\partial G_{0}}{\partial y}=\int dx\ \gamma _{0}^{2}\ K(x)\ \left[
G_{0}(y+\ln x)\ G_{0}(y+\ln (1-x))-G_{0}(y)\right]   \label{G0eq}
\end{equation}
\[
\frac{\partial G_{1}}{\partial y}=\int dx\ \gamma _{0}^{2}\ K(x)\ \left[
G_{0}(y+\ln x)\ \frac{G_{1}(y+\ln (1-x))}{(1-x)^{\alpha }}+\right.
\]
\qquad
\begin{equation}
\qquad +\left. G_{0}(y+\ln (1-x))\frac{G_{1}(y+\ln x)}{x^{\alpha }}%
-G_{1}(y)\right] \ +\frac{\partial G_{0}}{\partial y},  \label{G1eq}
\end{equation}
with the initial conditions $G_{0}(y=0)=u$ and $G_{1}(y=0)=0$. Let us note,
that physically the evolution equation in the parton branching angle $\theta
$ Eq.(\ref{General}) and the evolution equations in $y$ Eqs.(\ref{G0eq},\ref
{G1eq}) have very different meaning. The former is really following the
evolution of the intrajet distribution proceeding by emitting the gluons
into decreasing opening angles. The latter ones are describing the
evolution of our snapshot view of the cascade at a certain level. The
initial condition here is formulated not at the origin of the parton tree,
but at its very end. The condition $G_{0}(y=0)=u$ actually means, that a
parton having a minimal transverse momentum can not evolve at all. Larger $y$
mean that there is some phase space for degrading down to $Q_{0}$. In this
sense the evolution in $y$ goes in the opposite direction to the more
intuitive angular one. At the same time using $y$ as an evolution variable
is crucial in formulating a scale-invariant picture of the perturbative
cascade, so it will be this evolution that will be studied in the following.
Let us also note, that we shall restrict our consideration to
the simple case of pure gluodynamics and frozen coupling constant
 $\gamma_{0}=const$.

Before proceeding to solving Eqs.(\ref{G0eq},\ref{G1eq}) it is necessary to
understand what are the integration limits in these equations. The most
natural procedure is to demand the positivity of the arguments of the
generating functions entering Eqs.(\ref{G0eq},\ref{G1eq}). This is
equivalent to the requirement of stopping the evolution completely at $%
p_{t}=Q_{0}$ (i.e. no particle could be emitted with $p_t < Q_0$).
Then the integration over the energy fraction inherited by the
daughter parton is restricted to the interval
\begin{equation}
\int dx\ldots =\int\limits_{e^{-y}}^{1-e^{-y}}dx\dots   \label{limits}
\end{equation}
In the high energy limit of large $y$, where we expect our basic
perturbative treatment to hold, one can make a replacement
\begin{equation}
\int dx\dots =\int\limits_{0}^{1}dx\dots ,
\end{equation}
provided that the corresponding integrals are convergent.

In general from Eq.(\ref{limits}) it is clear, that this constraint becomes
increasingly important as we are moving towards the borderline of
perturbative  evolution , which corresponds to a starting point in the
evolution equation in $y$. Thus the accuracy of the predictions following
from the evolution equation Eq.(\ref{General}) is to a certain extent
limited by the necessity of fixing the initial conditions at the "least
perturbative" end of the intrajet evolution.

\section{Perturbative Contribution to Moments}

For reader's convenience we briefly remind in this Section some important
formulas in the solution \cite{DH} of the zeroth approximation Eq.(\ref{G0eq})
for the factorial moments of the intrajet particle distributions
in the frozen coupling constant case $\gamma _{0}=const$. Further details of
this solution can be found in \cite{DH}, \cite{D1}.

Let us start with computing the mean multiplicity

\begin{equation}
n_{0}(y)=\left. \frac{\partial G_{0}(y|u)}{\partial u}\right| _{u=1}
\end{equation}
From Eq.(\ref{G0eq}) and taking into account that $G_{0}(y|1)=1$ we get a
differential equation on mean multiplicity

\begin{equation}
\frac{\partial n_{0}}{\partial y}=\gamma _{0}^{2}\int dx K(x)\{n_{0}(y+\ln
(x))+n_{0}(y+\ln (1-x))-n_{0}(y)\}
\end{equation}
with an obvious initial condition $n_{0}(0)=1$. The solution is
\begin{equation}
n_{0}(y)=e^{\gamma y}
\end{equation}
where $\gamma $ is determined by a transcendent equation
\begin{equation}
\gamma =\gamma _{0}^{2}B(\gamma )  \label{gameq}
\end{equation}
\begin{equation}
B(a)=\int\limits_{0}^{1}dxK(x)\{x^{a}+(1-x)^{a}-1\},\quad a\neq 0,-1,-2,...
\end{equation}
The function $B(a)$ is plotted in Fig.1. The solution of Eq.(\ref{gameq})
determines $\gamma $ as a function of $\gamma _{0}^{2}$ (and therefore
as a function of $\alpha_s$). The function
$\gamma (\alpha_s)$ is plotted in Fig. 2.

It is well known \cite{D1} that in the high energy limit the KNO-scaling,
characterizing the intrajet evolution, is equivalent to the factorial moments
 F-scaling , so that in the expansion
\begin{equation}
G_{0}(y|u)=1+\sum\limits_{q=1}^{\infty }\frac{(u-1)^{q}}{q!}%
F_{q}^{(0)}n_{0}(y)^{q}=1+\sum\limits_{q=1}^{\infty }\frac{(u-1)^{q}}{q!}%
F_{q}^{(0)}e^{q\gamma y},\ \ y\rightarrow \infty   \label{Fq0}
\end{equation}
where $F_{q}^{(0)}$ are the factorial moments of the intrajet particle
distribution, these factorial moments do not depend on $y$. After
substituting the above expansion into Eq.(\ref{G0eq}) one obtains a
quadratic recurrent relation for the factorial moments
\begin{equation}  \label{F0sol}
F_{q}^{(0)}=\frac{\gamma _{0}^{2}}{\gamma q-\gamma _{0}^{2}B(\gamma q)}%
\sum\limits_{m=1}^{q-1}C_{q}^{m}F_{m}^{(0)}F_{q-m}^{(0)}R(\gamma m,\gamma
(q-m)),\quad q>1,  \label{Fq0rec}
\end{equation}
where
\[
R(a,b)=\int dxK(x)x^{a}(1-x)^{b}=B(a,b+2)+\frac{1}{2}B(a+2,b+2),
\]
$B(a,b)$ being the Euler beta-function. The resulting values of the
factorial moments are easily obtained by numerically solving Eq.(\ref{Fq0rec}%
). In Fig.~4 the values of $F_{q}$ for some $\alpha_s$ are shown.

\section{Mean multiplicity}

In this section we calculate the corrections to the mean multiplicity
in the first order in $\varepsilon =\beta /p$, so let us take $%
n=n_{0}-\varepsilon n_{1}$, where $n_{1}$ is a correction term
that has to be calculated from the correction to the generating functional in
Eq.~(\ref{resol})
\begin{equation}
n_{1}(y)=\left. \frac{\partial G_{1}(y|u)}{\partial u}\right| _{u=1}
\end{equation}
From Eq.(\ref{G1eq}) we get a differential equation on the correction to
mean multiplicity
\begin{equation}
\frac{\partial n_{1}}{\partial y}=\gamma _{0}^{2}\
\int_{e^{-y}}^{1-e^{-y}}dxK(x)\left\{ \frac{n_{1}(y+\ln x)}{x^{\alpha }}+%
\frac{n_{1}(y+\ln (1-x))}{(1-x)^{\alpha }}-n_{1}(y)\right\} +\gamma
e^{\gamma y}  \label{n1eq}
\end{equation}
The above equation is a linear inhomogeneous one, so its
solution is a sum of a general solution to a homogeneous equation and a
particular solution to the inhomogeneous one. The solution of the
homogeneous equation is proportional to $e^{\gamma _{1}y}$, where $\gamma
_{1}$ is a solution of
\begin{equation}
\gamma _{1}=\gamma _{0}^{2}B(\gamma _{1}-\alpha )  \label{gam1eq}
\end{equation}
In this case we have taken the integration limits to be $(0,1)\ $, which here
can be done without loosing accuracy. From
Eqs.(\ref{gameq},\ref{gam1eq}), and the fact that $B(a)$ is a
decreasing function of $a$, there follows an important inequality
\begin{equation}
\max (\gamma ,\alpha )<\gamma _{1}<\alpha +\gamma .  \label{neq}
\end{equation}
for $\alpha > 0$ so that the solution to the homogeneous equation is dominating
the high energy asymtotics $y\to \infty $. The function $\gamma $ is plotted
in Fig.2 as a function of $\alpha _{s}$ and in Fig.3 as a function of $\alpha $.
In the following we shall need only the high energy asymptotics of the
solution, which is provided by the solution of the homogeneous equation $%
e^{\gamma _{1}y}$:
\begin{equation}
n(y)=e^{\gamma y}-\left( \frac{\beta }{p}\right) ^{\alpha }C_{\gamma
_{1}}e^{\gamma _{1}y}=e^{\gamma y}\left( 1-C_{\gamma _{1}}\left( \frac{\beta
\theta }{Q_{0}}\right) ^{\alpha }e^{(\gamma _{1}-\gamma -\alpha )y}\right) .
\end{equation}
so the correction is known up to the prefactor which is determined from the
initial conditions and can be found only from the full solution of
Eq.(\ref{n1eq}). From the above equation and Eq.(\ref{neq}) we see that the
relative magnitude of the dissipative correction is enhanced with respect to
 a naively expected product of the factors $\sim e^{-\alpha y}$ and
$e^{-\gamma y}$, the latter obviously being a typical $1/n_0$ preasymptotic
 correction.

\section{Factorial moments.}

In this section we will calculate the nonperturbative corrections to the
factorial moments of the intrajet particle distribution. Let us remind (see,
e.g., \cite{BPQ} and \cite{D1}) that the factorial moments $F_{q}$ are
defined as coefficients in the expansion of the generating functional in a
power series in the mean multiplicity
\begin{equation}
G(p,\theta |u)=1+\sum\limits_{q=1}^{\infty }\frac{(u-1)^{q}}{q!}F_{q}n^{q},\
F_{1}=1.  \label{Mom}
\end{equation}
To the first order in the small parameter $\varepsilon$ we get from
Eq.~(\ref{Mom})
\[
G=1+\sum\limits_{q=1}^{\infty }\frac{F_{q}^{(0)}-\varepsilon
F_{q}^{(1)}}{q!}(n_{0}-\varepsilon n_{1})^{q}(u-1)^{q}=
\]

\begin{equation}
=1+\sum\limits_{q=1}^{\infty }\frac{F_{q}^{(0)}}{q!}n_{0}^{q}(u-1)^{q}-%
\varepsilon \sum\limits_{q=1}^{\infty }\frac{%
qF_{q}^{(0)}n_{1}+F_{q}^{(1)}n_{0}}{q!}n_{0}^{q-1}(u-1)^{q},  \label{Mom1}
\end{equation}
where $\varepsilon F_q^{(1)}$ is a dissipative correction to the $q$-th
factorial moment and we have explicitly isolated the nonperturbative correction
to the mean multiplicity calculated in the previous paragraph.
 From Eq.~(\ref{Mom1}) we immediately read off the expression for the
first order correction to the generating functional $G_1$:

\begin{equation}
G_{1}=\sum\limits_{q=1}^{\infty }\frac{qF_{q}^{(0)}n_{1}+F_{q}^{(1)}n_{0}}{q!%
}n_{0}^{q-1}(u-1)^{q}=\sum\limits_{q=1}^{\infty }\frac{\Phi _{q}(y)}{q!}%
(u-1)^{q}  \label{Phi-intro}
\end{equation}
where $\Phi _{q}(y) = (qF_{q}^{(0)}n_{1}+F_{q}^{(1)}n_{0})n_{0}^{q-1}$  and
the ''initial conditions" $F_{1}=F_{1}^{(0)}=1$ give $\Phi _{1}(y)=n_{1}(y)$%
. A complete calculation of $\Phi _{q}(y)$ is possible but very
cumbersome. To the accuracy we are interested in it will be sufficient to
work with the leading contribution to $\Phi _{q}(y)$, greatly simplifying
the calculations and making the results  transparent.

The full equation on $\Phi _{q}(y)$ reads

\[
\Phi _{q}^{\prime }(y)=\gamma _{0}^{2}\ \int dx\ K(x)\left\{ \frac{1}{%
x^{\alpha }}\Phi _{q}(y+\ln x)+\frac{1}{(1-x)^{\alpha }}\Phi _{q}(y+\ln
(1-x))-\Phi _{q}(y)\right\} +
\]

\begin{eqnarray}
&&+\gamma _{0}^{2}\ \sum\limits_{m=1}^{q-1}\ C_{q}^{m}F_{q-m}^{(0)}e^{\gamma
(q-m)y}\int dx\,K(x)\left\{ \frac{\Phi _{m}(y+\ln x)}{x^{\alpha }}%
(1-x)^{\gamma (q-m)}\right. +  \label{Phiq1} \\
&&+\left. \frac{\Phi _{m}(y+\ln (1-x))}{(1-x)^{\alpha }}x^{\gamma
(q-m)}\right\} +\gamma qF_{q}^{(0)}e^{q\gamma y}.  \nonumber
\end{eqnarray}
The homogeneous part of Eq.(\ref{Phiq1}) is the same as for in the equation
for $n_{1}$ Eq.(\ref{n1eq}), so its solution has the same form $%
C_{q}e^{\gamma _{1}y}$. The solution to the inhomogeneous equation on $\Phi
_{q}(y)$ is a sum of some exponents in $y$ (this immediately follows from
the linearity of Eq.(\ref{Phiq1}) and the structure of $\Phi _{q^{\prime }}$s
with $q^{\prime }<q$). Proceeding this way we get $\Phi _{q}\sim \Phi
_{m}e^{(q-m)\gamma y}$ ($m<q$), where the sign $\sim $ means that $\Phi _{q}$
includes the terms from $\Phi _{m}e^{(q-m)\gamma y}$. All such components
together with the solution of the homogeneous part of the equation give
$\Phi _{q}\sim e^{(q-m)\gamma y+\gamma _{1}y},e^{q\gamma y}\ (q\geq m\geq 1)$%
. From Eq.(\ref{neq}) it is clear that the leading term has the form $%
e^{(q-1)\gamma y+\gamma _{1}y}$. Let us define
\begin{equation}
\Phi _{q}(y)=\Psi _{q}\ e^{((q-1)\gamma +\gamma _{1})y}=\Psi _{q}\ e^{\gamma
_{q}y},\ \mbox{where}\ \gamma _{q}=(q-1)\gamma +\gamma _{1}.
\label{Psiqsubs}
\end{equation}
Substituting this Ansatz to Eq.(\ref{Phiq1}) we have
\[
\gamma _{q}\Psi _{q}=\gamma _{0}^{2}\ \sum\limits_{m=1}^{q-1}\
C_{q}^{m}F_{q-m}^{(0)}\Psi _{m}\int dx\,K(x)\left\{ x^{\gamma _{m}-\alpha
}(1-x)^{\gamma (q-m)}+(1-x)^{\gamma _{m}-\alpha }x^{\gamma (q-m)}\right\} +
\]
\begin{equation}
+\gamma _{0}^{2}\Psi _{q}\int dx\,K(x)\left\{ x^{\gamma _{q}-\alpha
}+(1-x)^{\gamma _{q}-\alpha }-1\right\} \,\,,\,\ q>1  \label{Psiqeq}
\end{equation}
$\Psi _{1}=C_{\gamma _{1}}.$ From Eq.(\ref{Psiqeq}) we immediately get a
recurrent relation on $\Psi _{q}$:
\begin{equation}
\Psi _{q}=\frac{\gamma _{0}^{2}\ }{\gamma _{q}-\gamma _{0}^{2}B(\gamma
_{q}-\alpha )}\sum\limits_{m=1}^{q-1}\ C_{q}^{m}\Psi _{m}F_{q-m}^{(0)}\left[
R(\gamma _{m}-\alpha ,\gamma (q-m))+R(\gamma (q-m),\gamma _{m}-\alpha
)\right] .  \label{Psiqres}
\end{equation}
Using Eqs.(\ref{Phi-intro}) and (\ref{Psiqsubs}) one can extract $F_{q}^{(1)}
$ from $\Psi _{q}$:
\[
qF_{q}^{(0)}\Psi _{1}e^{\gamma _{q}y}+F_{q}^{(1)}e^{q\gamma y}=\Psi
_{q}e^{\gamma _{q}y}\Rightarrow
\]
\begin{equation}
F_{q}^{(1)}=\left( \Psi _{q}-qF_{q}^{(0)}\Psi _{1}\right) e^{(\gamma
_{1}-\gamma )y},\ q\geq 1
\end{equation}
(for $q=1$ it is obvious that $\ F_{q}^{(1)}=0$). Let us further introduce a
convenient notation $\varepsilon _{q}$ by
\begin{equation}
\varepsilon _{q}=\frac{\Psi _{q}/\Psi _{1}}{F_{q}^{(0)}}-q,  \label{epsil1}
\end{equation}
so that in the zeroth and first orders in $\varepsilon =\left( \frac{\beta }{p}%
\right) ^{\alpha }$ we get
\begin{equation}
G=1+\sum\limits_{q=1}^{\infty }\frac{F_{q}^{(0)}(1-\varepsilon _{q}\left(
\frac{\beta \theta }{Q_{0}}\right) ^{\alpha }\Psi _{1}e^{(\gamma _{1}-\alpha
-\gamma )y})}{q!}n^{q}(u-1)^{q}  \label{epsil2}
\end{equation}
The function $\varepsilon _{q}$ is plotted in Fig.5.

In order to compare our final result with the existing experimental data on
the factorial moments it is convenient to introduce another function $\rho_q
$
\begin{equation}
\rho _{q}=%
{\displaystyle {1-F_{q}/F_{q}^{(0)} \over \varepsilon _{q}}}%
=\left( \frac{\beta \theta }{Q_{0}}\right) ^{\alpha }\Psi _{1}e^{(\gamma
_{1}-\alpha -\gamma )y}  \label{result}
\end{equation}
From this formula we see that the convenience of using $\rho _{q}$ is that
it directly measures the nonperturbative contribution to the factorial
moments. From Eq.(\ref{result}) we see, that our formalism predicts that $%
\rho_q$ is positive and does not depend on $q$.  Some plots of $\rho _{q}$
are shown in Figs. 6 and 7.

Let us now turn to the analysis of the existing experimental data on
factorial moments of the intrajet particle distributions. From Eq.(\ref
{result}) it is clear that our main prediction is that if we substitute into
Eq.(\ref{result}) the experimentally measured factorial moments $F_{q}^{exp}$
and compute the quantity
\begin{equation}
\eta ^{exp}=(1-F_{q}^{exp}/F_{q}^{(0)})=\rho _{q}\varepsilon _{q},
\end{equation}
the difference between the experimental and purely perturbative factorial
moments as measured by $\rho _{q}^{exp}$ should (a) be positive and (b) have
no dependence on the moment's rank $q$.

We have computed the function $\rho _{q}^{exp}$ for the experimental data on
intrajet factorial moments from
\cite{OPAL}. We see,
that $\rho _{q}^{exp}$ is indeed positive thus signalling the presence of
the contributions beyond the standard perturbative ones in the experimental
data. In our treatment it is tempting to assume, that this difference is due
to the influence of the soft nonperturbative modes on the process of
intrajet's evolution. From Fig. 6 we see, however, that the data show a clear
dependence of $\rho _{q}^{exp}$ on $q$. This varies somewhat with varying $%
\alpha $, but always stays quite pronounced. This shows that our treatment
of the nonperturbative effects is actually incomplete. However, this is
something that had to be expected, because here the small parameter
is actually $q\gamma $ \cite{D3}, so for large enough $q$ the correction
is already much
larger than the ''leading'' perturbative term. Therefore with our accuracy
we can meaningfully calculate only the corrections to the two first moments,
and here the independence of $q$ as a starting approximation does not seem
unreasonable. Let us also mention, that in order to get a complete
understanding of the possible nonperturbative contributions, it is desirable to
analyze the corresponding contributions brought in by conventional
hadronization physics in the Monte-Carlo approach (actually the local
parton-hadron duality we use seems to work quite well, see, e.g., the recent
review \cite{KO}) and by taking into account the corrections originated by the
running coupling.  Finally, let us look at the sensitivity of our results to
the choice of the value of the frozen perturbative coupling $\alpha_s$ and the
parameter $\alpha$ parametrizing the coupling to the nonperturbative modes.
In Fig. 6 we show ratio $\rho_q$ for 3 different values of $\alpha_s$ and
in Fig. 7 we show the dependence of the ratio $\rho_q$ on the value of
the parameter $\alpha$ fixing the functional form of the dissipative
interaction.  We see, that both dependencies are reasonably smooth.

\section{Conclusions.}

By using the previously developed formalism \cite{LO} we have computed a
nonperturbative preasymptotic contribution to the factorial moments of the
intrajet
particle distribution violating the perturbative scaling. The presence
of the additional nonperturbative contribution is supported by the existing
experimental data on the factorial moments of the intrajet particle
distributions. To arrive at a final conclusion on the magnitude of the
nonperturbative contribution one also has to take into account the
contributions from hadronization stage (conventionally parametrized by some
standard MC scheme), as well as those due to the running of the coupling
constant.

\bigskip

{\it Acknowledgments.}
We are grateful to I.M.~Dremin and I.V.~Andreev for reading the manuscript
and useful discussions. This work was supported by the Russian Foundation
for  Basic Research under Grant N 96-02-16210.

\newpage

\begin{center}
{\large {\bf Figure captions}}
\end{center}

\begin{itemize}
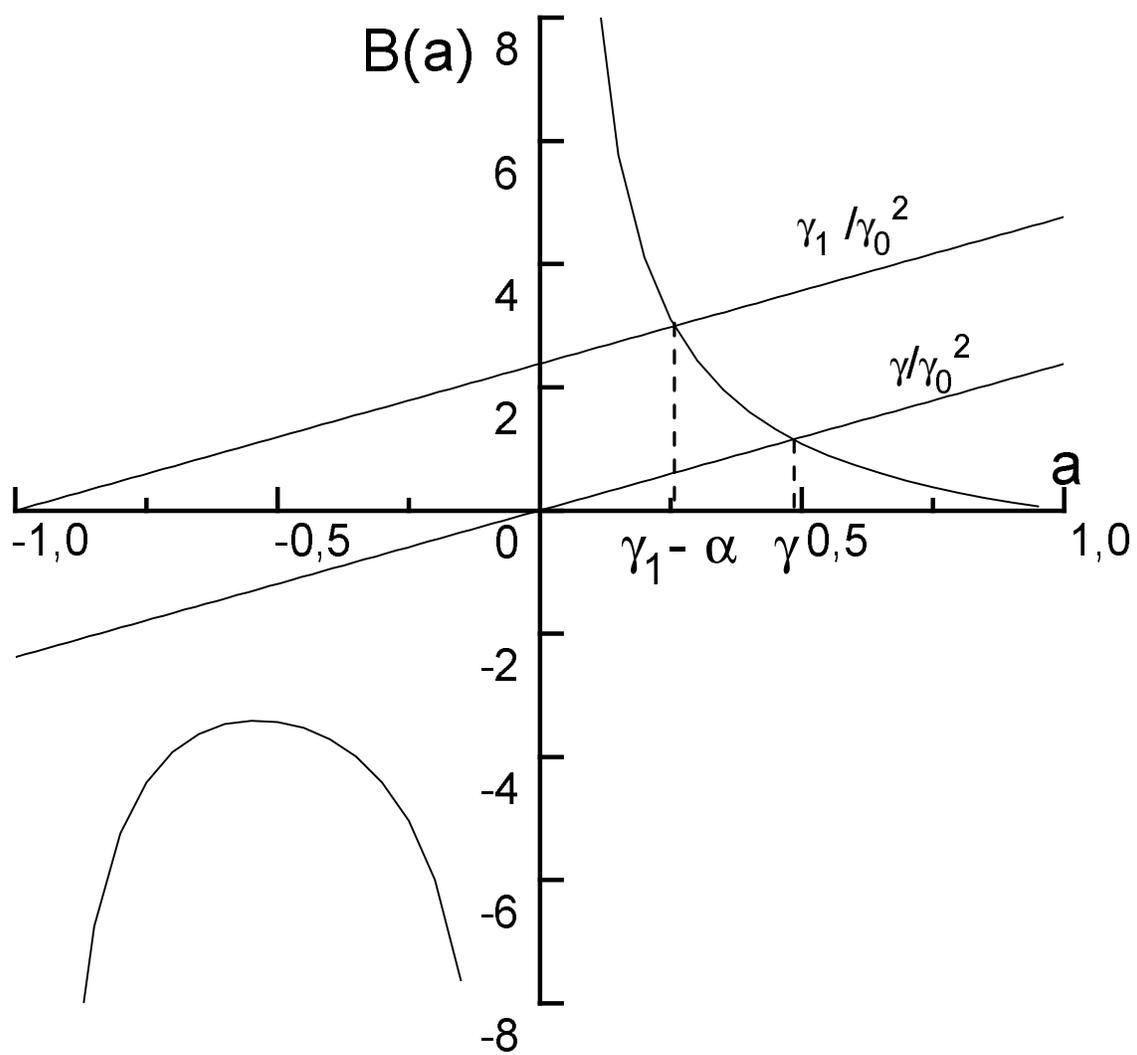
\item[{\bf Fig.1}]  The function $B(a)$  and a graphical calculation of
 $\gamma $ and $%
\gamma _{1}$ (see Eqs.(\ref{gameq}, \ref{gam1eq})) for $\alpha
_{s}=0.22\,(\gamma _{0}^{2}=0.42)$ and $\alpha =1$.

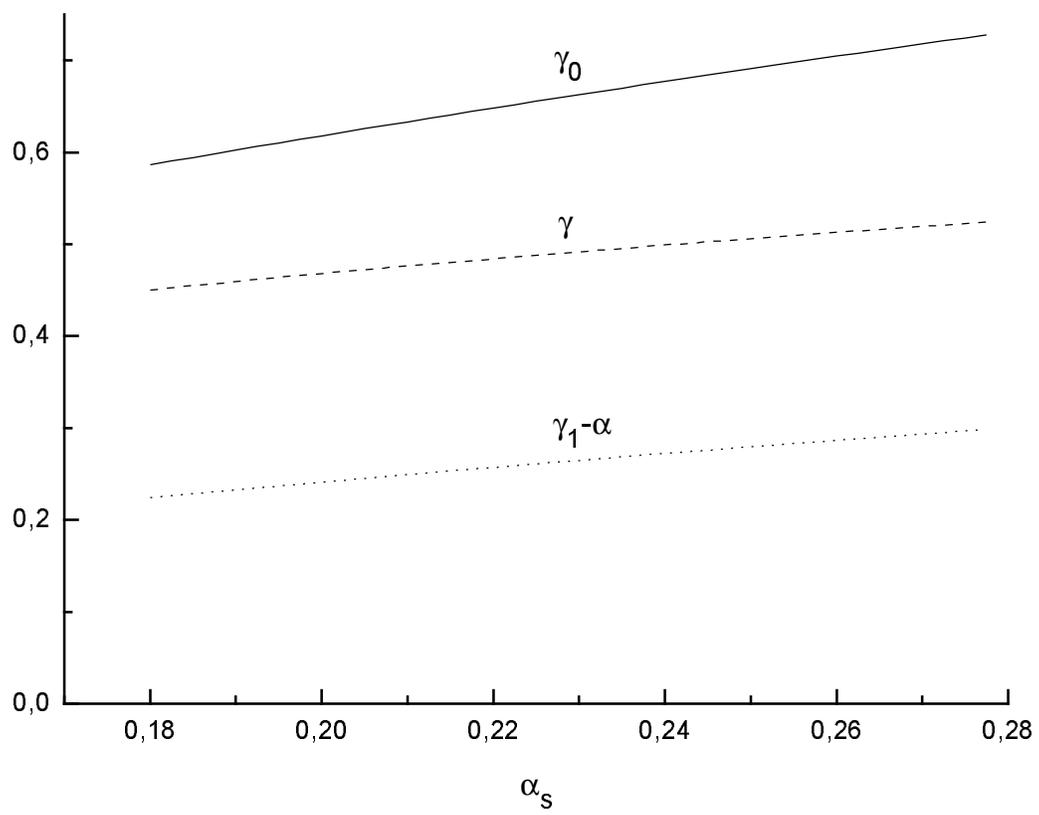
\item[{\bf Fig.2}]  $\gamma _{0}$, $\gamma $ and $\gamma _{1}-\alpha $ with
respect to $\alpha _{s}$ obtained from Eqs.(\ref{gameq},\ref{gam1eq}).

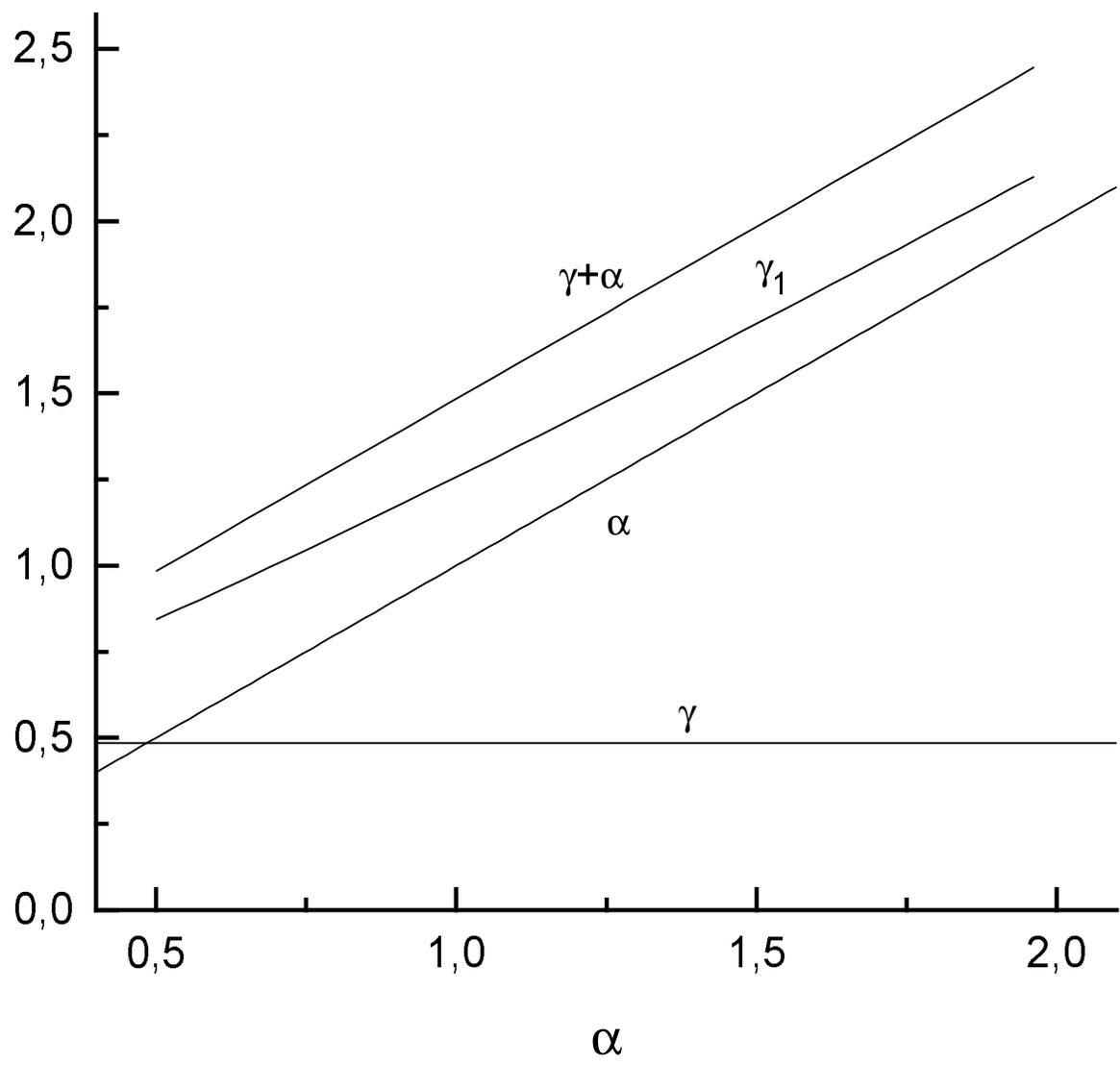
\item[{\bf Fig.3}]  $\gamma _{1}$ dependence on $\alpha $. See Eqs.(\ref
{gam1eq}, \ref{neq})

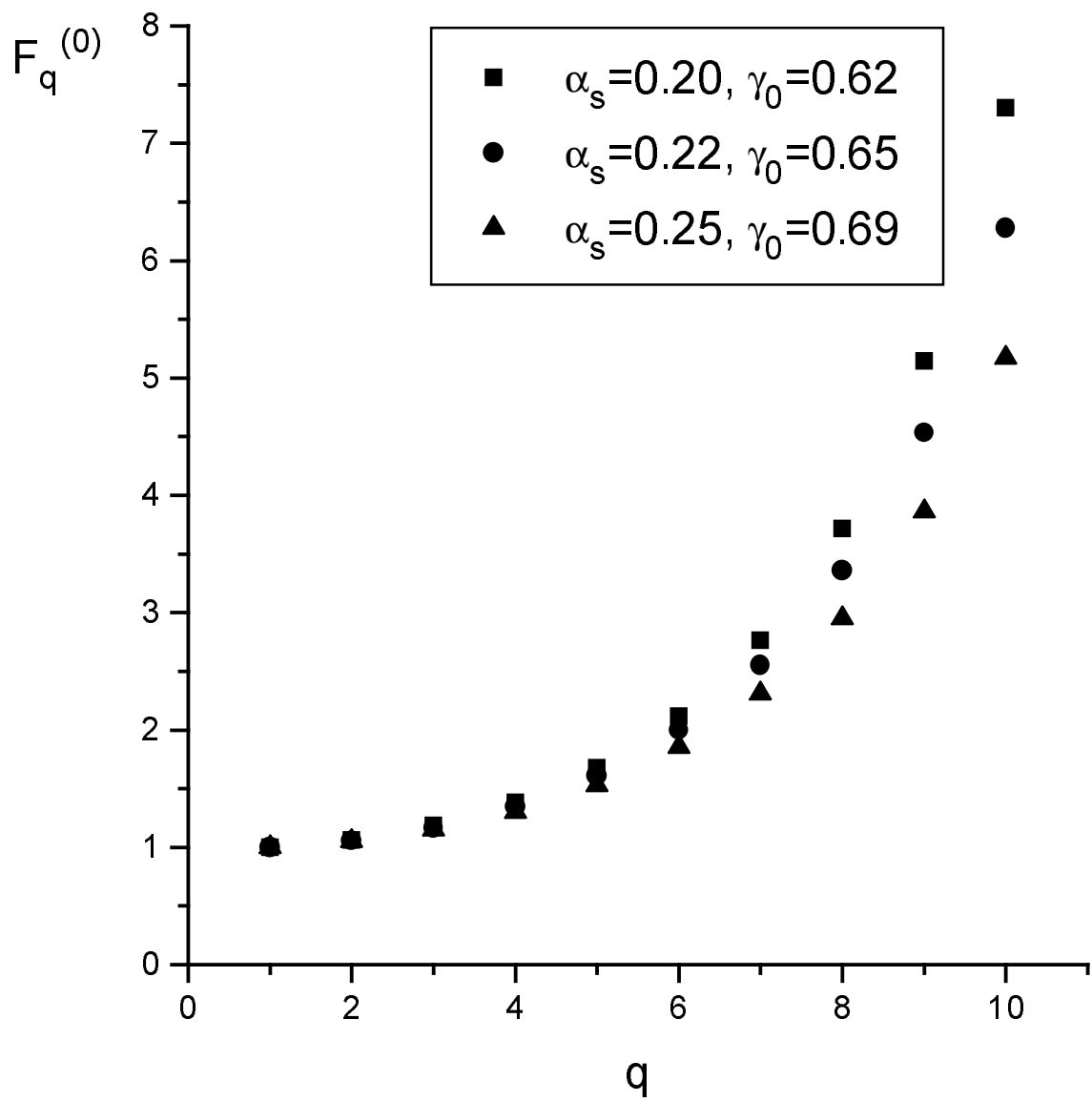
\item[{\bf Fig.4}]  $F_{q}^{(0)}$ for three values of $\alpha _{s}$. See Eq.(%
\ref{F0sol})

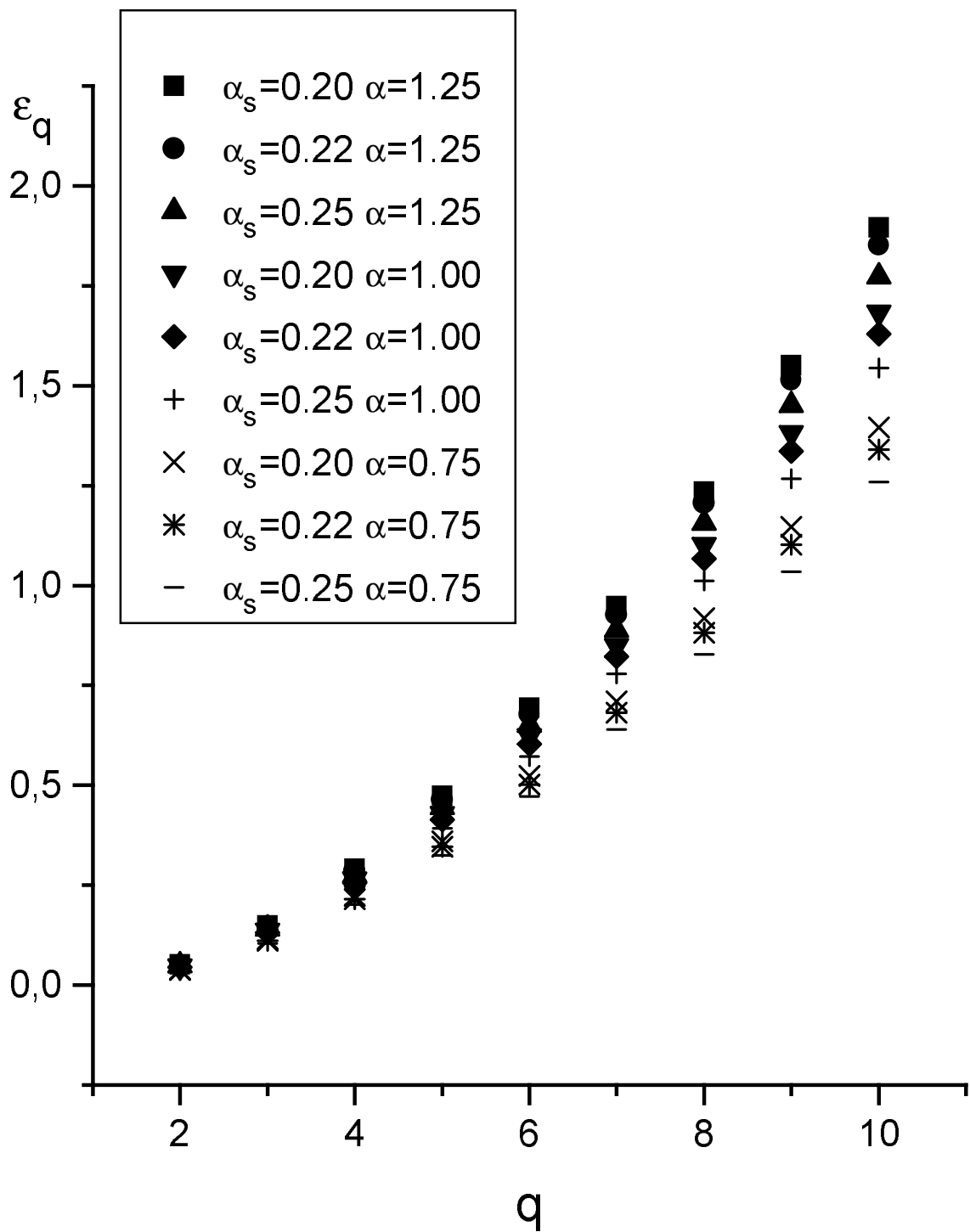
\item[{\bf Fig.5}]  $\varepsilon _{q}$ for different $\alpha _{s}$ and $%
\alpha $. See Eqs.(\ref{epsil1},\ref{epsil2})

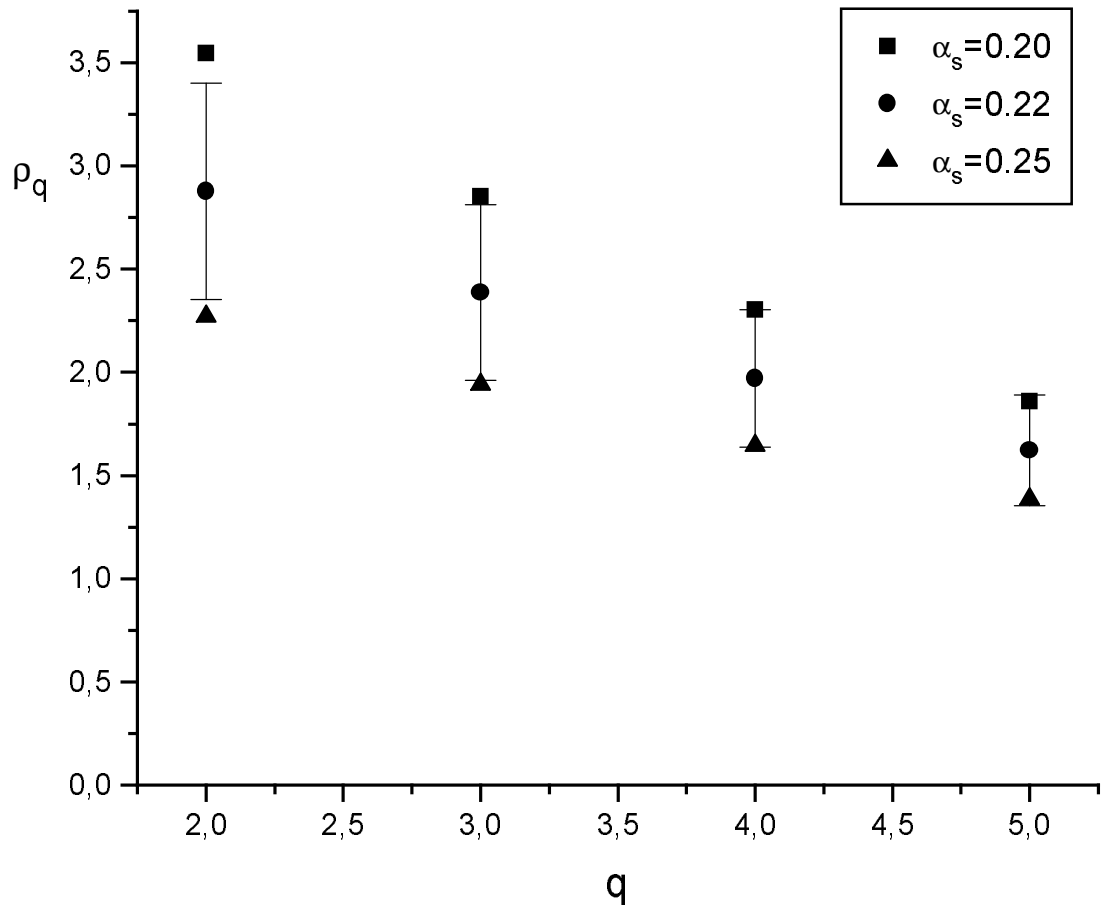
\item[{\bf Fig.6}]  $\rho _{q}$ dependence on $\alpha _{s}$, $\alpha =1$.
Based on OPAL\cite{OPAL} data for gluon jets.

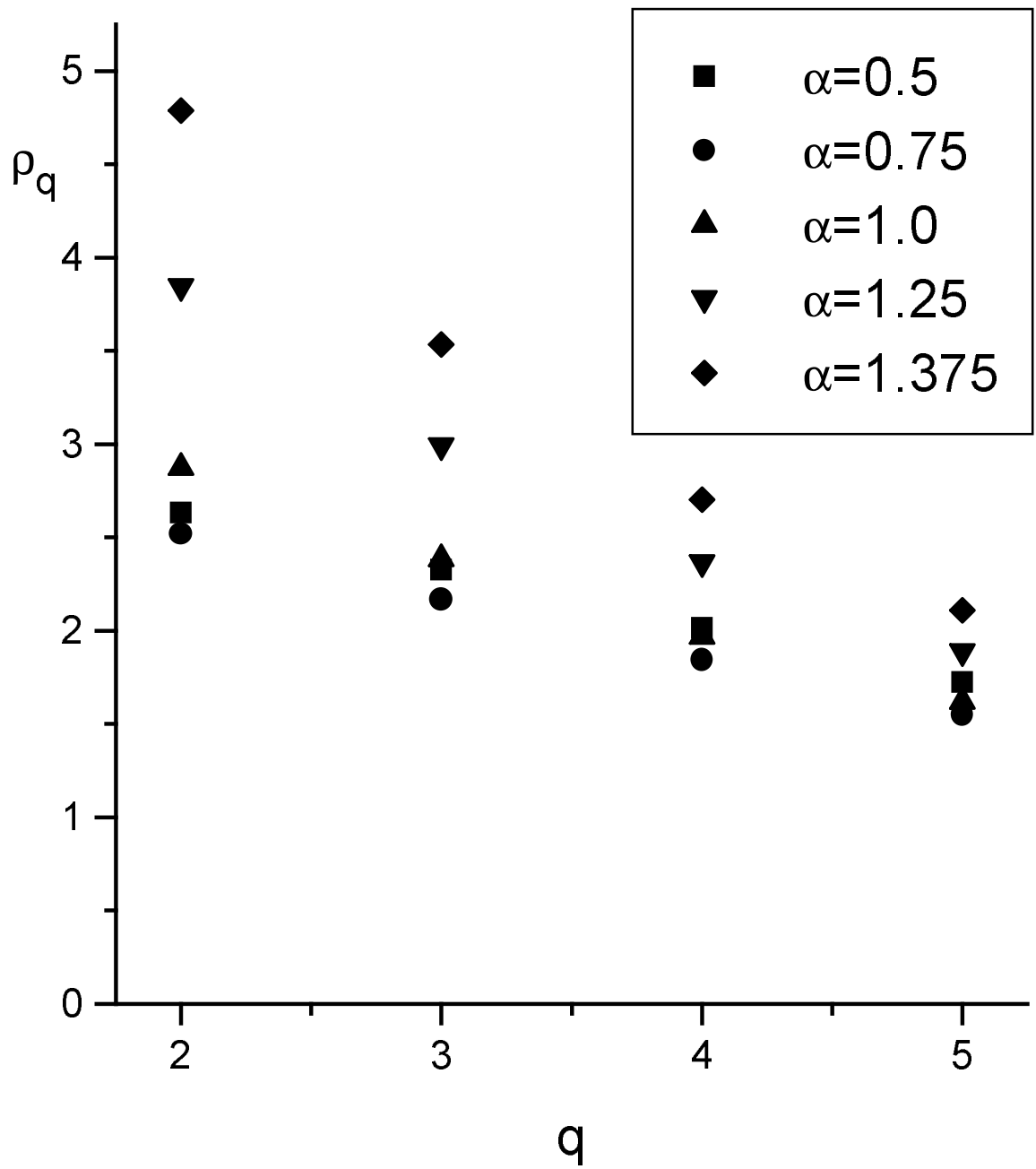
\item[{\bf Fig.7}]  $\rho _{q}$ dependence on $\alpha $ for $\alpha _{s}=0.22
$ Based on OPAL\cite{OPAL} data for gluon jets.
\end{itemize}

\end{document}